\documentclass[runningheads]{llncs}
\usepackage{graphicx}
\usepackage{hyperref}
\usepackage{xcolor}
\usepackage{wrapfig}
\usepackage{subfig}
\usepackage{longtable}
\usepackage{booktabs}
\usepackage{array}     
\usepackage{ragged2e}
\usepackage{float}

\usepackage{pifont}

\newcommand\blfootnote[1]{%
  \begingroup
  \renewcommand\thefootnote{}\footnote{#1}%
  \addtocounter{footnote}{-1}%
  \endgroup
}

\begin{document}
\title{Automatic Generation of Chatbots for Conversational Web Browsing}
\titlerunning{Conversational Web Browsing}
%
\author{Pietro Chitt\`o\inst{1} \and Marcos Baez\inst{2} \and Florian Daniel\inst{1} \and Boualem Benatallah\inst{3}}
\authorrunning{P. Chitt\`o et al.}
%
\institute{Politecnico di Milano,
Via Ponzio 34/5, 20133 Milan, Italy\\
\email{pietro.chitto@mail.polimi.it, florian.daniel@polimi.it}
\and
LIRIS – University of Claude Bernard Lyon 1, Villeurbanne, France \\
\email{marcos.baez@liris.cnrs.fr}
\and
University of News South Wales,
Sydney, Australia \\
\email{boualem@cse.unsw.edu.au}
}
\maketitle 
\begin{abstract}
In this paper, we describe the foundations for generating a chatbot out of a website equipped with simple, bot-specific HTML annotations. The approach is part of what we call \textit{conversational web browsing}, i.e., a dialog-based, natural language interaction with websites. The goal is to enable users to use content and functionality accessible through rendered UIs by ``talking to websites'' instead of by operating the graphical UI using keyboard and mouse.  The chatbot mediates between the user and the website, operates its graphical UI on behalf of the user, and informs the user about the state of interaction. We describe the conceptual vocabulary and annotation format, the supporting conversational middleware and techniques, and the implementation of a demo able to deliver conversational web browsing experiences through Amazon Alexa.


\end{abstract}
%
%


\section{Introduction}

Conversational agents\blfootnote{This is a post-peer-review, pre-copyedit version of an article accepted to the 29th International Conference on Conceptual Modeling, ER 2020.} are emerging as an exciting new platform for accessing online services that promise a more natural and accessible interaction paradigm. They have shown great potential for regular users in hands-free and eyes-free scenarios but also for making services more accessible to people with disabilities and visual impairments \cite{pradhan2018accessibility},  as well as groups, such as older adults, often challenged by service design choices \cite{kowalski2019older}. This new generation of agents is however not able to natively access the Web, requiring web developers and content creators to implement specific ``skills" to offer their content and services on Amazon Alexa, Google Assistant and other platforms. This requirement represents a huge barrier for developers and creators who might not have the skills or resources to invest, and a missed opportunity for making the Web accessible to everyone.

Integrating conversational capabilities into software enabled services is an emerging research topic \cite{baez2020patterns}, as pushed by recent works by Castaldo et al. \cite{castaldo2019conversational} on inferring bots directly from  database schemas, Yaghoub-Zadeh-Fard et al. \cite{yaghoub2020rest2bot} on deriving bots from APIs, and by Ripa et al. \cite{ripa2019end} on generating informational bots out of website content. While these works are facilitating chatbot integration at different levels of the Web architecture, they do not address the challenges of generating chatbots from both \textit{content} and \textit{functionality} available in websites. 

In this paper, we take a software engineering approach and study how to enable conversational browsing of websites equipped with purposefully designed annotations. This represents the first step towards our vision \cite{BaezConversations2019} of enabling users to access the content and services accessible through rendered UIs by ``talking to websites'' instead of by operating the graphical UI using keyboard and mouse. We start with an annotation-driven approach as the focus is to lay the foundation for conversational browsing and to identify all necessary conversational features and technical solutions, which can then lead to the development of support tools and automatic approaches. In doing so, we make the following contributions:

\begin{itemize}
     \item conceptual vocabulary for augmenting websites with conversational capabilities, able to describe domain knowledge (content and functionality) while abstracting interaction knowledge (enacting low-level interactions with sites); 
     \item an approach, architecture and techniques for generating a chatbot out of a website equipped with simple, bot-specific HTML annotations;
     \item prototype implementation and technical feasibility of the proposed automatic chatbot generation approach.
 \end{itemize}

In the following we describe a concrete target scenario, the overall approach, and the prototype implementation.

\section{Scenario and Requirements}

\begin{figure}[t!]
\centering
  \includegraphics[width=\columnwidth]{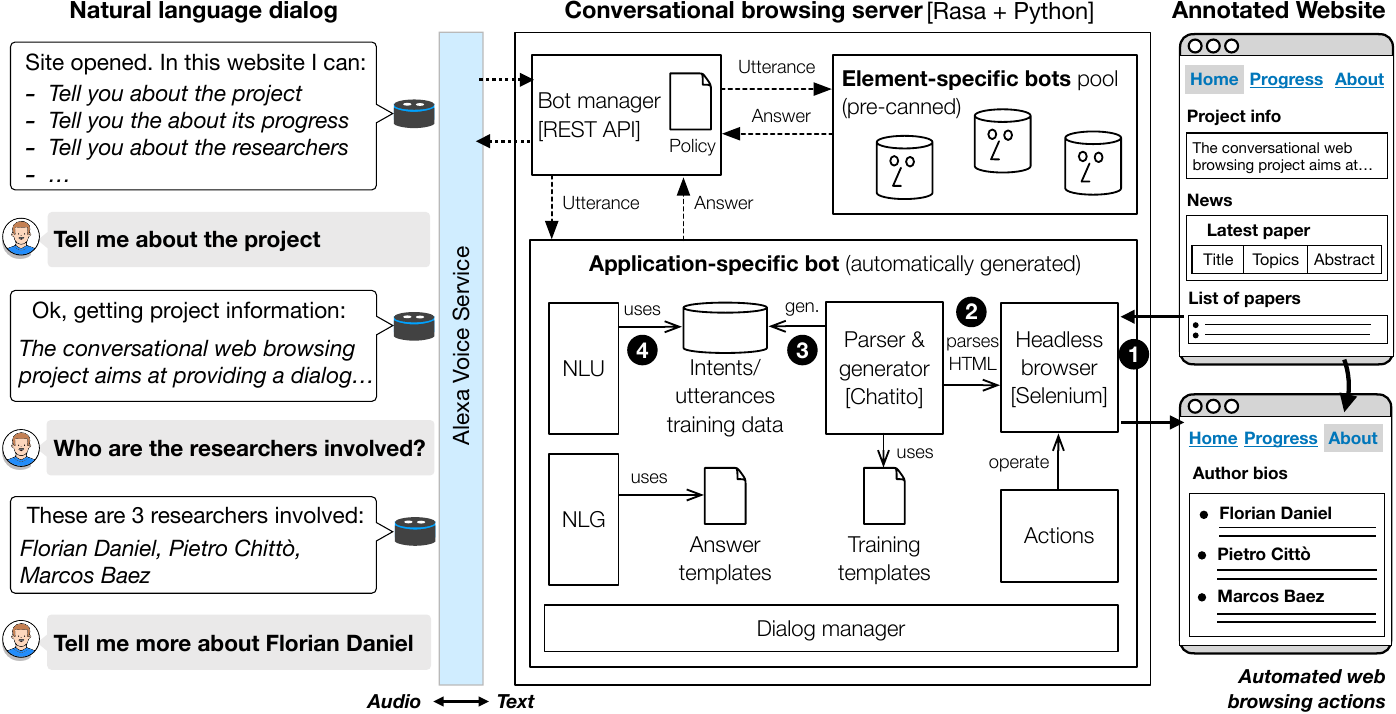}
  \caption{Conversational browsing scenario: the user talks to a bot not the website. }
  \label{fig:example}
\end{figure}

We describe our target scenario by illustrating the interactions of a user browsing a typical research project website using a smart speaker such as  Amazon Echo (Figure \ref{fig:example}). 
After the user requests access to the research project website, a conversational agent tailored to the website content, functionality and domain knowledge is automatically generated to mediate the interactions between the user and the target website. During these interactions, i) the user is informed of the available features, ii) can browse the website in dialog-based natural language interactions with the agent, and iii) the agent identifies and performs the appropriate web browsing actions on the target website on behalf of the user

Before diving into the requirements posed by the envisioned scenario, we need to introduce some concepts related to chatbot development, in what refers to \textit{task-oriented} chatbots.
Modern task-oriented chatbots are built on a frame-based architecture, which relies on a domain ontology (composed of frame, slots and values) that specify the type of user intentions the system can recognize and respond to \cite{jurafsky2017dialog}. 
\textit{Intents} refer to the task requested by the user and the \textit{actions} to the specific operations performed by the chatbot to serve the intent.  Identifying user intents given a \textit{request} in natural language (e.g., ``Tell me about Florian Daniel”) requires a natural language processing component \textit{trained} with a dataset of examples (e.g., researcher\_info: [``Tell me about @researcher", ``Who is @researcher?", ...]) to correctly classify the request and infer the slots and values (e.g., intent: researcher\_info, researcher : ``Florian Daniel"). Then the \textit{dialog management} component, based on intent, the input provided and the conversation context, decides on the appropriate \textit{action} (e.g., parse associated DOM element). A \textit{response} is generated using a natural language generation component that elaborates the results and presents them in a format that fits the conversation medium (refer to \cite{jurafsky2017dialog} for more on chatbot design and architecture). 

Having introduced the scenario and main concepts, we refine some key requirements to enabling conversational browsing as identified earlier \cite{BaezConversations2019}:

\begin{itemize}

\item[R1]   \textbf{Orientation}: The bot must be able to summarize the content and/or functionalities offered by the website, to guide users through site offers at any point and to provide for basic access structures (e.g., ``In this site you can..."). 

\item[R2]  \textbf{Inferring intents and parameters}: The bot must be able to understand the user's intent and enact suitable actions in response. Intents may be \textit{application-agnostic} (e.g., fill a form field) or \textit{application-specific} (e.g., post a new paper). The latter requires the bot to infer the intents from the website. 

\item[R3] \textbf{Training and vocabulary}: The bot should be able to speak and understand the language of the target website, so as to identify intents and elaborate proper responses. This requires deriving domain knowledge directly from the website, training the bot to identify application-specific intents.

\item[R4] \textbf{Browsing actions enactment}: As the bot mediates between the user and the website, enacting an action in response to an identified intent requires a strategy for translating high-level user requests into automated low level interactions with the website.

\item[R5] \textbf{Dialog control from rendered UIs}. As the user browse the website conversationally, the chatbot should track the state of the dialog and choose dialog actions considering  the evolving  state of the rendered UI. That is, it should consider the conversation context as well as the browsing context.

\end{itemize}

\section{Conversational Web Browsing: Approach}
\label{sec:browsing}

The approach illustrated in Figure \ref{fig:example} is based on three main ingredients (i) purposefully designed bot \textit{annotations}, (ii) a \textit{middleware} comprised of chatbot generation and run-time units, and iii) a medium-specific \textit{conversational interface}. 
Web developers enable conversational access by augmenting their websites with bot-specific \textbf{annotations}, which associate knowledge about how to generate a conversational agent with specific HTML constructs. 
Initiating a conversational browsing session then triggers the chatbot \textbf{generation process}. This process is about generating an application-specific bot tailored to the intents and domain knowledge of the target website, while reusing a library of generic element-specific bots.
Using a conversational interface (e.g., Amazon Echo) the user can start a dialog with the website. At \textbf{run-time}, the middleware processes the user requests in natural language, selects the relevant bot and executes the appropriate actions on the rendered GUI of the website.

Supporting conversational browsing is not trivial and requires weighing several options. The most important decisions that resulted in our solution are:

\begin{itemize}

\item \textbf{Domain vs. interaction knowledge}: Using a website generally requires the user to master two types of knowledge, \textit{domain knowledge} (to understand content and functionalities) and \textit{interaction knowledge} (to use and operate the site). 
This distinction is powerful to separate concerns in conversational browsing. Domain knowledge, e.g., about the research project and scientific publications, must be provided by the developer, as this varies from site to site.
Interaction knowledge, e.g., how to fill a form or read text paragraph by paragraph, can be pre-canned and reused across multiple sites. We thus distinguish between an \textit{application-specific} bot and a set of \textit{element-specific} bots [R1,R2]. The former masters the domain, the latter enable the user to interact with specific content elements like lists, text, tables, forms, etc.

\item \textbf{Modularization}: Incidentally, the distinction between application- and ele\-ment-specific bots represents an excellent opportunity for modularization and reuse. Application-specific bots must be generated for each site anew [R3]; element-specific bots can be implemented and \textit{trained once and reused multiple times}. They can be implemented for specific HTML elements, such as a form, or they can be implemented for a very specific version thereof, e.g., a login form. 
However, the presence of application- and element-specific bots introduces the need for a suitable bot selection logic.

\item \textbf{Bot selection}: 
As a user may provide as input any possible utterance at any instant of time, referring to either application-specific or element-specific intents, it is not possible to pre-define conversational paths through a website. Instead, some form of random access must be supported. We introduce for this purpose a so-called \textsf{bot manager}, which takes as input the utterance and forwards it to the bots registered in the system [R5]. Depending on the context (e.g., the last used bot) and the confidence provided by each invoked bot, it then decides which bot is most likely to provide the correct answer [R1,R2]. 
Thanks to the \textsf{bot manager}, the ensemble of application-specific and element-specific bots presents itself as one single bot to the user.

\end{itemize}

\section{Annotating Websites with Conversational Knowledge}
\label{sec:annotation}

The goal of the work presented in this paper is to prevent asking developers to provide full-fledged chatbots for their websites in order to support conversational browsing.
The challenge is asking them to provide as little information as possible -- the annotation -- such that, together with the content and functionality that are already in the site (its GUI), it is possible to automatically generate a chatbot. 

\smallskip
\noindent \textbf{Conceptual model.} Let's start with introducing the key concepts that enable conversational browsing.
Figure \ref{fig:concept} uses an intuitive, graphical notation to contextualize them in a model of a simple website about a research project, e.g., our project on conversational browsing. The site consists of a set of pages, of which the model ignores the actual content; the design of such content has traditionally been approached by modeling languages like WebML \cite{CFBBCM02} or IFML \cite{ifml}. Instead, the model hypothesizes a conceptual vocabulary that could extend the pages, subsuming the presence of suitable content\footnote{Note that here we do not want to introduce an own, new modeling notation for conversational browsing; Figure \ref{fig:concept} serves an intuitive, illustrative purpose only.}. We identified these concepts through a literature and systems review and prototyping efforts:

\begin{figure}[t!]
\centering
  \includegraphics[width=.9\columnwidth]{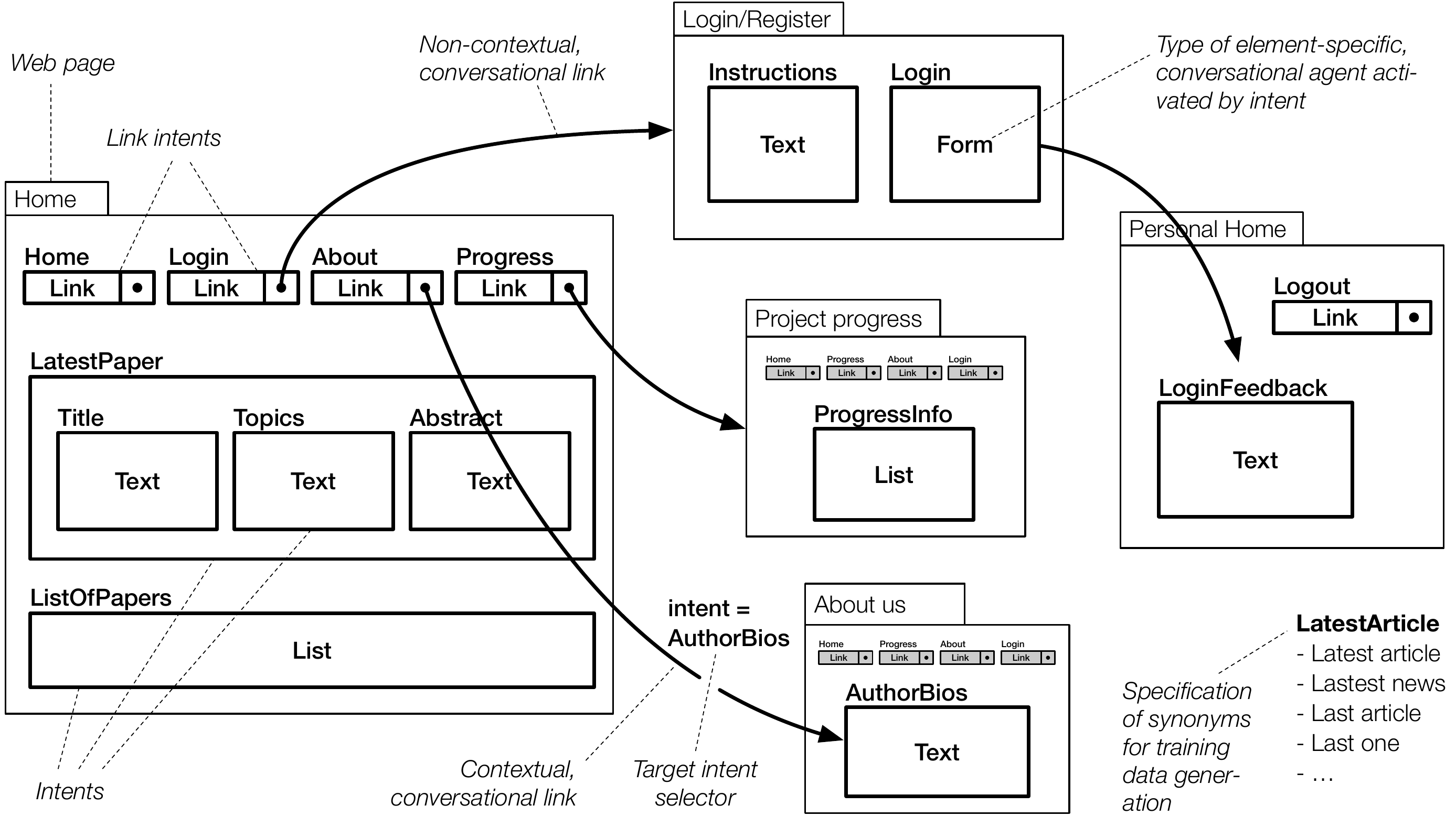}
  \caption{Informal graphical model of a project website explaining the core concepts of application-specific, conversational browsing.
  Labels in italics define the used graphical notation. Gray-shaded intents are copied from the \textsf{Home} page.}
  \label{fig:concept}
\end{figure}

\begin{itemize}

\item \textbf{Intents}: These are the core ingredients of conversational browsing. 
Intents annotate HTML constructs and thereby qualify their contents as relevant for the enactment of the intents' actions [R2]. More importantly, intents enable the user to access content and functionality. We distinguish three types:

    \begin{itemize}
    \item \textbf{Selection intents} identify HTML constructs the developer wants to make accessible through the chatbot. In order to guide user inside complex pages, selection intents can be structured hierarchically, which tells the bot to read out options at different levels of detail. 
    \item \textbf{Link intents} enable the user to navigate among pages of the site. 
    Each navigation may reset the context of the conversation and prompt the bot to inform the user of the new intents available. 
    \item \textbf{Built-in intents} are the intents that the framework comes with in order to support basic interactions, such as orienting the user inside a page by proactively telling him/her which options are available (e.g., ``What is the page about?")[R1]. Built-in intents do not require any annotation.
    \end{itemize}
    
\item \textbf{Conversational links}: These are the counterpart of hyperlinks in conversational browsing and tell link intents their target [R4]. Similar to conventional links, we distinguish two types of conversational links:

    \begin{itemize}

    \item \textbf{Non-contextual} conversational links are links that can be navigated with the help from the bot and result in the loading and rendering of a new page, causing the bot to start a new browsing context. 
    That is, each page accessed through a non-contextual link causes the bot to inform the user about the content of the page [R1]. For example, \textsf{Login} follows a non-contextual link to a new page (with a different menu of options), triggering the bot to inform of the available options (Instructions, Login).     
    \item \textbf{Contextual} conversational links are links that are directed not only toward a new page but also toward a specific target intent. 
    If a user thus accesses a page through a contextual link, the bot will immediately start performing the action associated with the target intent [R5], e.g.,  \textsf{About} (contextual link) will trigger \textsf{AboutBios} (reading the associated text).    
    
    \end{itemize}

\item \textbf{Bot types}: If a selection intent identifies the HTML construct to act upon, i.e., if it cannot be further split into sub-intents (e.g., \textsf{LatestPaper}  $\rightarrow$ \textsf{Title}), the type of element-specific bot able to perform the expected action can be specified (\textsf{Title}: Text). As explained earlier, the number of element-specific bots is theoretically unlimited, but we identify the need for a minimum set of element-specific bots able to manage the following content elements [R2]: 

    \begin{itemize}
    \item \textbf{Text}, i.e., text organized into headings, sub-headings and paragraphs. Element-specific actions are reading out loud the full text, reading the titles only, jumping back and forth among paragraphs, etc.
    \item \textbf{List}, i.e., an ordered or unordered list of items. Element-specific actions 
    are telling the number of items, reading them out, navigating them, etc.
    \item \textbf{Table}, i.e., content organized in rows and columns. Element-specific actions are reading 
    by cells, navigating by rows, reading by column, etc.
    \item \textbf{Form}, i.e., input fields grouped together and accompanied by a submission button. Element-specific actions are telling which inputs are required, filling individual fields, confirming inputs, submitting, etc.
    \end{itemize}

\item \textbf{Domain vocabulary}: It is necessary to equip all intents in the website with their domain-specific vocabulary. This can be achieved by accompanying intents with \textit{labels} and \textit{synonyms} that can be used to generate combinations of phrases and to train the application-specific bot [R3]. For instance, the intent \textsf{LatestPaper} 
with the words
\textsf{``latest paper, recent paper''} or similar. 

\item \textbf{Intent description}:
Intent descriptions are simple textual explanations that the bot can use to tell the user which intents a given page supports. For instance, the \textsf{LatestPaper} intent could be described using the words \textsf{``tell you about the last paper published by the project''} [R1]. 

\end{itemize}

Given a website, it is important to note how the sensible selection of which HTML construct to annotate and how to connect them with conversational links allows the developer to construct pre-defined \textbf{dialog flows} guiding the user through the content and functionalities published by a website [R5].

\begin{figure}[t!]
\centering
  \includegraphics[width=\columnwidth]{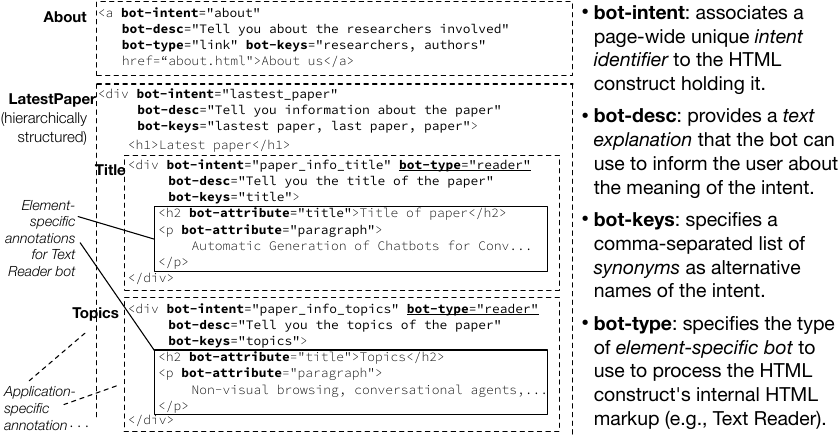}
  \caption{Simplified code excerpt of the \textsf{$<$body$>$} of the \textsf{Home} page in Figure \ref{fig:concept} with annotations for conversational browsing. Application-specific annotations enable navigation and content access; element-specific ones instruct the \textsf{Text Reader}.}
  \label{fig:annotation}
\end{figure}

\smallskip
\noindent \textbf{Annotation format}.
\textit{Annotating} a website now means associating conversational knowledge (knowledge about how to generate a conversational agent) with specific HTML constructs in a page. 
The \textit{cues} for the generation of the agent come in the form of HTML attributes and developer-provided values. Informed by the conceptual model, the concrete attributes for the generation of \textbf{application-specific} bots are highlighted in Figure \ref{fig:annotation}.
The figure provides a practical example of the use of these attributes, and the use of one \textbf{element-specific} attribute: \textbf{\textsf{bot-attribute}}, which identifies element-specific \textit{content types} that the respective element-specific bot can understand.
While some annotations may seem redundant (e.g., can be derived from HTML tags), developers not always follow the semantics of HTML tags. For instance, one of the most used tags today is the \textsf{$<$div$>$} tag, which lacks semantics. Explicit annotations can also allow developers indicate what elements to expose to the chatbot.

As research progresses, we intend to maintain an up-to-date version of the annotation format on \textbf{GitHub} and to improve it with the help from the community. Please refer to \url{https://github.com/floriandanielit/conversationalweb}.

\section{Generating Application-Specific Conversational Agent}
\label{sec:generation}

The generation process can be divided into two phases: 
(i) the generation of the \textit{application-specific training data} and the training of the \textsf{NLU} (natural language understanding), and (ii) the generation of a suitable \textit{conversational context model} to enable the \textsf{bot manager} to manage the dialog.
The generation of the application-specific training data follows the steps highlighted in Figure \ref{fig:example} using circled numbers: the \textsf{headless browser} loads the current page of the website and builds its DOM \ding{202}, the \textsf{parser and generator} extracts intent identifiers and the list of intent synonyms \ding{203} and generates a dataset of utterances for training \ding{204}; the \textsf{NLU} uses the dataset to learn intents and application-specific vocabulary \ding{205}.

The \textbf{conversational context model} is generated by the \textsf{parser and generator} once the \textsf{NLU} is successfully trained. It consists in a tree representation\footnote{The tree is a result of the hierarchical organization enabled by selection intents, e.g., \textsf{LatestPaper}$\rightarrow$\textsf{Title}(Text Reader)}
of the intents contained in the current page: $CT = \langle N, C \rangle$, where $N$ is the set of nodes, where each node represents one application-specific intent in the page, and $C = N \times N$ represents the set of non-cyclic, directed child relationships of the tree.
Each node $n \in N, n = \langle intent, type, desc, keys, elem, link \rangle$ contains the identifier, type, description and keywords of the respective intent, the HTML element it is associated with, and the possible conversational link in case the intent is a link intent. The root node $r \in N$ represents the information intent associated with the \textsf{$<$body$>$} element of the current page. Intermediate nodes represent access intents with sub-intents; leaf nodes (nodes without children) represent intents to be processed using a given $type$ of element-specific bot.

The \textsf{bot manager} now uses the so constructed context model to decide which bot to choose to advance the conversation with the user. The proposed \textsf{policy} works as follows:  as the user provides input, the \textsf{bot manager} checks if the last used bot (the \textit{current} bot) is able to understand the input, i.e., if it is able to identify an intent with a confidence that exceeds a given threshold $\tau$. If yes, the respective answer is forwarded to the user, otherwise it forwards the input to all direct children of the current bot, and recursively to the sub-children if none is successful. If any of them is able to identify an intent with sufficient confidence, that bot becomes the new current bot and its answer is forwarded to the user. If the current bot corresponds to a leaf node and is not able to understand the user input, it escalates the input to upper levels until there is a higher-level bot able to understand the input or the escalation reaches the root node. If none is able understand the input, the user is asked to reformulate his/her request.

\section{System Implementation and Technical Validation}

The conversational browsing infrastructure outlined in Figure \ref{fig:example} has been implemented making use of ready technologies: 
\textit{Alexa Voice Service} 
for voice to text conversion,
\textit{Rasa NLU} (\url{https://rasa.com/}) for natural language understanding,
\textit{Selenium} (\url{https://selenium.dev/}) as headless browser integrated with \textit{Mozilla Firefox}, 
and 
\textit{Chatito} (\url{https://github.com/rodrigopivi/Chatito}) for the generation of training data. Custom integration and chatbot code were written in \textit{Python}. 
For the tests with Alexa, the infrastructure was deployed on Heroku.

While the training phase of the chatbot could be done once of the entire site, in our current prototype  we opted for a page-by-page training, in order to support dynamically generated pages.
As the focus of the prototype was technical feasibility, it is not yet optimized for performance. However, tests on a local machine (Omen by HP 15-DH0, Intel Core i7, 16 GB of RAM, SSD hard-drive, Win10 64bit) show that page loading and rendering, training data generation and bot training requires up to few seconds, an acceptable performance for some scenarios. Fetching pages from the Web adds an additional overhead. The construction of the context model is negligible in terms of execution time.

The element-specific bots of the prototype are custom Rasa bots with pre-defined intents, actions and NLU models. Demo videos illustrating the components of the approach can be found at 
\url{https://bit.ly/2OckzZW}.

\section{Related Work}

The problem of \textbf{non visual web browsing} has produced two main approaches: \textit{markup-based} approaches such as VoiceXML \cite{oshry2007voice} and \textit{voice-enabled screen readers} integrated into web browsers \cite{ashok2015capti}. VoiceXML \cite{oshry2007voice} is a W3C markup language for voice applications typically accessed using a phone. Applications are stand alone and could complement websites, but there is no native integration of the two. Voice-based screen readers (e.g., \cite{ashok2015capti}) aim at lowering the complexity of managing shortcuts in navigating with screen readers, enabling users to utter browsing commands in natural language (``press the cart button”). 
While  valuable, these approaches were developed to support desktop web browsing: they require users to be aware of the layout of the pages and perform low-level, step-by-step interactions, or to create macros to automate tasks. 

As for \textbf{chatbot development}, general platforms and tools support the development of stand-alone chatbots (e.g., DialogFlow, Instabot.io). 
Another approach is that of deriving chatbots directly from database schemas, API definitions and web content. Prominent works in this regard are the ones by Castaldo et al. \cite{castaldo2019conversational} exploring the idea of conversational data exploration, by inferring a chatbot directly from annotated database schema; 
Yaghoub-Zadeh-Fard et al. \cite{yaghoub2020rest2bot} generating a conversational interface directly from API specifications (e.g., OpenAPI).
Website content has also been used for chatbot generation. Popular in e-commerce and CRM, approaches such as SuperAgent \cite{cui2017superagent} can generate conversational FAQ based on the content to visitors directly on the website. Ripa et al. \cite{ripa2019end} focus on making informational queries over content intensive websites accessible via voice-based interfaces (e.g., smart speakers), relying on augmentations provided by end-users. While all these works illustrate the diversity of approaches, they require either (bot) programming knowledge (and effort), are constrained by an application domain, or are limited to Q\&A.

\section{Conclusion and Outlook}
This paper contributes with abstractions, techniques and conceptual vocabulary for superimposing conversational bots over websites. These contributions along with the software infrastructure enable the (semi)automatic generation of chatbots directly from websites, and can be leveraged by authoring tools to enable developers, even without chatbot skills, to obtain chatbots effectively and efficiently.
The solution presented is a proof-of-concept implementation not optimized for large applications, and thus presents points for improvement that are the focus of our ongoing work. As a next step, we will out user studies with different types of target users (end users and developers) and derive guidelines for conversational browsing.
We are also already studying how to use machine learning and AI along with existing Web technical specifications (e.g., HTML5) to replace some explicit annotations by automatic recognition.

\bibliographystyle{splncs04}
\bibliography{bot}

\end{document}